\documentclass[aps,pra,amsmath,amssymb,amsfonts,twocolumn,superscriptaddress,a4paper,10pt]{revtex4-2}
\usepackage{amsmath,amssymb,amsthm} 
\usepackage{physics}
\usepackage{graphicx}
\usepackage{hyperref}        
\usepackage{bm}              
\usepackage{lipsum}          
\usepackage{newtxtext}
\usepackage{newtxmath}
\usepackage{tikz}
\usetikzlibrary{
    positioning,
    arrows.meta,
    chains
}
\hypersetup{
    colorlinks=true,
    linkcolor=blue,
    citecolor=blue,
    urlcolor=blue
}

\makeatletter

\def\twocolumngridnorule{%
 \clearpage
 \@twocolumntrue
 \col@number\tw@
}%

\makeatother

\theoremstyle{remark}
\newtheorem*{theorem*}{Theorem}

\theoremstyle{remark}
\newtheorem*{claim*}{Claim}

\begin{document}


\title{Network Nonlocality Sharing in Generalized Star Network from Bipartite Bell Inequalities}

\author{Hao-Miao Jiang}
\author{Xiang-Jiang Chen}
\author{Liu-Jun Wang}
\author{Qing Chen}
\email{chenqing@ynu.edu.cn}
\affiliation{School of Physics and Astronomy and Yunnan Key Laboratory for
	Quantum Information, Yunnan University, Kunming 650500, China}

\date{\today} 

\begin{abstract}

This work investigates network nonlocality sharing for a broad class of bipartite Bell inequalities in a generalized star network with an 
$(n,m,k)$ configuration, comprising $n$ independent branches, $m$ sequential Alices per branch, and $k$ measurement settings per party. On each branch, the intermediate Alices implement optimal weak measurements, whereas the final Alice and the central Bob perform sharp projective measurements. Network nonlocality sharing is witnessed when the quantum values of the network correlations associated with relevant parties simultaneously violate a star-network Bell inequality generated from the given class of bipartite Bell inequalities. We streamline the calculation of the quantum values of the network correlations and derive an analytical expression for the bipartite quantum correlator, valid for arbitrary measurement settings and weak-measurement strengths. The network nonlocality sharing for Vértesi inequalities has been studied within the framework, and simultaneous violations are found in $(2,2,6)$ and $(2,2,465)$ cases, with the latter exhibiting greater robustness. Our approach suggests a practical route to studying network nonlocality sharing by utilizing diverse bipartite Bell inequalities beyond the commonly used CHSH-type constructions.

\end{abstract}

\maketitle 


\section{Introduction}
\label{sec:intro}
Quantum nonlocality was first placed on a rigorous footing by Bell's theorem in 1964, which showed that the statistical predictions of quantum mechanics are incompatible with any theory based on local hidden variables (LHVs)~\cite{bellEinsteinPodolskyRosen1964}. In operational terms, this incompatibility is encoded in Bell inequalities: linear constraints satisfied by all local models but violated by certain quantum correlations. Bell’s work thus transformed earlier debates on the completeness of quantum mechanics into experimentally testable criteria for nonlocality and established Bell inequalities as fundamental tools for identifying and quantifying nonclassical correlations that are now viewed as a resource in quantum information (for a review see~\cite{brunnerBellNonlocality2014}).

Beyond the standard Bell scenario with a single source and a small number of parties, network nonlocality has emerged as an active area of research (see Ref.~\cite{tavakoliBellNonlocalityNetworks2022a} for a review). In most network scenarios, several independent sources distribute physical systems to multiple nodes, and classical descriptions are typically formulated under both locality and explicit source-independence assumptions. These modeling assumptions lead to new kinds of nonlocal correlations that cannot be fully captured by standard Bell inequalities and instead require distinct families of network Bell inequalities~\cite{skrzypczykEmergenceQuantumCorrelations2009,branciardCharacterizingNonlocalCorrelations2010a,branciardBilocalNonbilocalCorrelations2012}. A variety of network topologies have been investigated (see~\cite{fritzBellsTheoremCorrelation2012} for an overview on correlations with different topologies), including bilocal chains~\cite{skrzypczykEmergenceQuantumCorrelations2009,branciardCharacterizingNonlocalCorrelations2010a,branciardBilocalNonbilocalCorrelations2012,branciardClassicalSimulationEntanglement2012,gisinAllEntangledPure2017,tavakoliBilocalBellInequalities2021a}, star networks~\cite{tavakoliNonlocalCorrelationsStarnetwork2014,andreoliMaximalQubitViolation2017a,tavakoliCorrelationsStarNetworks2017a,munshiGeneralized$n$localityInequalities2021a}, triangle networks~\cite{gisinElegantJointQuantum2017,gisinEntanglement25Years2019a,renouGenuineQuantumNonlocality2019a,mukherjeeDetectingNontrilocalCorrelations2022a}, and other network architectures~\cite{rossetNonlinearBellInequalities2016a,yangNonlocalCorrelationsTreetensornetwork2021a,renouNetworkNonlocalityRigidity2022,renouNonlocalityGenericNetworks2022,sasmalNonlocalCorrelationsAsymmetric2023a}, revealing rich structures of network-induced nonlocality and stimulating ongoing work on both foundational questions and potential applications in quantum networks.

Nonlocality sharing, first demonstrated by Silva \textit{et al.}~\cite{silvaMultipleObserversCan2015}, addresses whether a single entangled pair can generate Bell-inequality violations for several observers who access one subsystem sequentially. In the standard sharing scenario, a sequence of observers performs weak measurements on the same subsystem of an entangled pair, each seeking to demonstrate nonlocality with a distant partner via violation of the CHSH inequality. By carefully tuning the measurement strength, nonlocal correlations can be shared among several observers without being destroyed by earlier measurements. Subsequent theoretical and experimental studies have substantially developed the sharing of Bell nonlocality~\cite{malSharingNonlocalitySingle2016a,dasFacetsBipartiteNonlocality2019b,renPassiveActiveNonlocality2019b,fengObservationNonlocalitySharing2020b,brownArbitrarilyManyIndependent2020a,chengLimitationsSharingBell2021a,steffinlongoProjectiveMeasurementsAre2022b,renNonlocalitySharingThreequbit2022a,yangSharingQuantumNonlocality2023a,schiavonThreeobserverBellInequality2017b,huObservationNonlocalitySharing2018b,xiaoExperimentalSharingBell2024a}. Sharing of other kinds of correlations has also been investigated, including steering~\cite{sasmalSteeringSingleSystem2018a,guptaGenuineEinsteinPodolskyRosenSteering2021a,yaoSteeringSharingTwoqubit2021a,hanActivationEinsteinPodolsky2024a,choiDemonstrationSimultaneousQuantum2020a,liuWitnessingMultiobserverSteering2022a,zhuEinsteinPodolskyRosenSteeringTwosided2022a}, entanglement~\cite{srivastavaEntanglementWitnessingArbitrarily2022a,huSequentialSharingTwoqudit2023a,liSequentiallyWitnessingEntanglement2024a,folettoExperimentalCertificationSustained2020a}, contextuality~\cite{kumariSharingNonlocalityNontrivial2019a,anwerNoiserobustPreparationContextuality2021a,chaturvediCharacterisingBoundingSet2021,kumariSharingPreparationContextuality2023a}, and network nonlocality~\cite{houNetworkNonlocalitySharing2022,wangNetworkNonlocalitySharing2022,dasResourcetheoreticEfficacySingle2022a,mahatoSharingNonlocalityQuantum2022b,halderLimitsNetworkNonlocality2022b,maoRecyclingNonlocalityQuantum2023b,kumarSharingNonlocalityNetwork2023b,zhangSharingQuantumNonlocality2023b,sunNetworkNonlocalitySharing2024b,caiFullNetworkNonlocality2024b}(for reviews see Refs.~\cite{photonics10121314,caiReviewQuantumCorrelation2025a}).

Here we focus on the network nonlocality sharing in a generalized star network. The generalized star network, also referred to as the \((n,m,k)\)-star network, comprises one central observer and \(n\) branches, with \(m\) observers on each branch and \(k\) measurement settings available to every observer. The first study of network nonlocality sharing~\cite{houNetworkNonlocalitySharing2022} considered a minimal configuration that can be regarded as a \((2,2,2)\)-star network, where sharing is witnessed by the simultaneous violation of the Branciard–Rosset–Gisin–Pironio (BRGP) inequalities~\cite{branciardBilocalNonbilocalCorrelations2012}. Network nonlocality sharing based on \(n\)-branch star inequalities~\cite{tavakoliNonlocalCorrelationsStarnetwork2014} has also been demonstrated experimentally in a \((3,2,2)\)-star network~\cite{maoRecyclingNonlocalityQuantum2023b}. Both BRGP and the $n$-branch star inequalities can be viewed as star-network inequalities derived from the CHSH inequality via the mapping introduced in~\cite{tavakoliCorrelationsStarNetworks2017a}. Subsequent work~\cite{wangNetworkNonlocalitySharing2022} studied network nonlocality sharing constructed from generalized CHSH inequalities~\cite{wehnerTsirelsonBoundsGeneralized2006}, allowing both the number of branches and the number of measurement settings to vary arbitrarily in an \((n,m,k)\) configuration. In their scenario, however, the sharing effect vanishes for \(k>3\), which motivates the search for explicit network constructions in which nonlocality sharing persists even when the number of measurement settings is large.

In this work, we develop a general sharing scheme for $n$-local star-network inequalities derived from a broad family of bipartite Bell inequalities. Our construction recovers earlier network sharing constructions based on generalized CHSH inequalities~\cite{wangNetworkNonlocalitySharing2022} as a special case. Central to our approach is an analytical expression for the bipartite correlator between Bob and any given Alice on a branch, assuming all intermediate Alices perform optimal weak measurements. This expression streamlines the evaluation of quantum network correlations and renders the numerical analysis of large-k configurations computationally feasible. As an application, we compute the network quantum correlations based on Vértesi inequalities~\cite{vertesiMoreEfficientBell2008} in the $(2,2,3)$, $(2,2,6)$, and $(2,2,465)$ configurations. We observe simultaneous violations for $k=6$ and $k=465$, with the $k=465$ case exhibiting greater sharing robustness.

This paper is organized as follows. In Sec.~\ref{sec:tgowssnandvsni}, we introduce the generalized star network, formulate the corresponding $n$-LHV model, and present the construction of $(n,m,k)$-star network inequalities from a broad family of bipartite Bell inequalities while preserving their classical bounds. In Sec.~\ref{sec:quantumviol}, we derive a general analytic expression for the bipartite quantum correlator in the optimal weak sharing generalized star network and show how it can be used to compute the associated star-network correlations, including previously studied generalized CHSH scenarios as special cases. In Sec.~\ref{sec:EGvertesi}, we introduce the sharing scheme on the $(n,m,k)$-star network and apply the scheme to the Vértesi inequalities. Finally, Sec.~\ref{sec:summary} summarizes our main results and outlines several open directions.

\section{The generalized star network \texorpdfstring{$n$}{n}-locality Scenario}\label{sec:tgowssnandvsni}

\subsection{LHV Model for The Generalized Star Network}\label{sec:nmkstarnetworklocality}

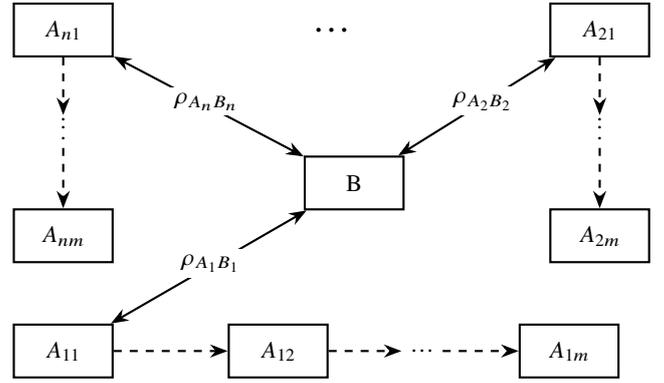
\begin{figure}[htbp]
\centering
\begin{tikzpicture}[
    node_style/.style={
        rectangle,
        draw,
        thick,
        minimum width=1.3cm,
        minimum height=0.7cm,
        align=center
    },
    arrow_style/.style={
        Stealth-Stealth,
        thick
    },
    dashed_arrow_style/.style={
        -Stealth,
        thick,
        dashed
    },
    label_style/.style={
        midway,
        fill=white,
        inner sep=2pt,
        font=\small 
    }
]


\node[node_style] (B) at (0,0) {B};

\node[node_style] (A11) [below left = 1.5cm and 2.5cm of B] {$A_{11}$};
\node[node_style] (A12) [right = 1.5cm of A11] {$A_{12}$};
\node[node_style] (A1m) [right = 2.5cm of A12] {$A_{1m}$};
\node[node_style] (Anm) [above = 0.8cm of A11] {$A_{nm}$};

\node[node_style] (An1) [above = 2cm of Anm] {$A_{n1}$};
\node (top_dots) [right = 2.5cm of An1] {\Large \dots};
\node[node_style] (A21) [right = 2.5cm of top_dots] {$A_{21}$};

\node[node_style] (A2m) [below = 2cm of A21] {$A_{2m}$};


\node (bot_dots) [left = 1cm of A1m] {\dots};
\node (lef_dots) [above = 0.8cm of Anm] {\vdots};
\node (ldcoo) [fill=white, inner sep=1pt] at (lef_dots) {.};
\node (rit_dots) [above = 0.8cm of A2m] {\vdots};
\node (rdcoo) [fill=white, inner sep=1pt] at (rit_dots) {.};


\path[arrow_style] (B) edge node[label_style, above, pos=0.6, xshift=8pt] {$\rho_{A_1 B_1}$} (A11);
\path[arrow_style] (B) edge node[label_style, above, pos=0.4, xshift=8pt] {$\rho_{A_2 B_2}$} (A21);
\path[arrow_style] (An1) edge node[label_style, above, pos=0.6, xshift=-8pt] {$\rho_{A_n B_n}$} (B);

\path[dashed_arrow_style] (A11) edge (A12);
\path[dashed_arrow_style] (A12) edge (bot_dots);
\path[dashed_arrow_style] (bot_dots) edge (A1m);
\path[dashed_arrow_style] (An1) edge (ldcoo);
\path[dashed_arrow_style] (lef_dots) edge (Anm);
\path[dashed_arrow_style] (A21) edge (rdcoo);
\path[dashed_arrow_style] (rit_dots) edge (A2m);
\end{tikzpicture}
\vspace{1.5em}
\caption{Schematic of the generalized star network. A central node $B$ shares $n$ bipartite states $\rho_{A_i B_i}$ with the first observers $A_{i1}$ ($i=1,\dots,n$), thereby generating $n$ branches. Along each branch $i$, the state is passed sequentially through $m$ parties $A_{i1},A_{i2},\dots,A_{im}$: every intermediate Alice performs an optimal weak measurement and forwards the post-measurement state to the next party, while the last Alice $A_{im}$ performs a sharp measurement. Each Alice has $k$ possible measurement settings. Double-headed arrows represent the distribution of the bipartite states $\rho_{A_i B_i}$ between $B$ and $A_{i1}$, and dashed arrows represent the sequential propagation of the post-measurement states along each branch.}
\label{fig:star_network}
\end{figure}

We first introduce the generalized star network scenario schematically illustrated in Fig.~\ref{fig:star_network}. The network consists of $n$ branches radiating from a central node (Bob). On each branch $i\in\{1,\dots,n\}$, there is a chain of $m$ parties, denoted by Alices $A_{i1},A_{i2},\dots,A_{im}$, arranged in a fixed order along that branch. Each Alice has access to $k$ possible measurement settings. 
The measurement procedure is as follows. The central Bob performs a single strong measurement on each of the $n$ entangled states that he initially shares with the first Alices $A_{11},\dots,A_{n1}$, one per branch. Once the post-measurement state is received by $A_{i1}$ on branch $i$, she performs an optimal weak measurement (as fully discussed in~\cite{silvaMultipleObserversCan2015}) and forwards the resulting post-measurement state to the next Alice $A_{i2}$ on the same branch. This sequential weak-measurement process is repeated by each intermediate party $A_{i j^{(i)}}$, when $j^{(i)}\in\{1,\dots,m-1\}$, along branch $i$, until the final party $A_{im}$, who performs a sharp measurement and does not forward the state further.
\vspace{0.3em}

Given this star-shaped and sequential structure, and assuming that the sources on different branches are independent, an $n$-local probability distribution compatible with an LHV model reads
\begin{gather}
    P(a_{1 j^{(1)}},\dots,a_{n j^{(n)}},b_1,\dots,b_n|x_{1 j^{(1)}},\dots,x_{n j^{(n)}},y)\nonumber\\ =\int \left[ \prod_{i=1}^n \mathrm{d}\lambda_i q_i(\lambda_i)\,p(a_{i j^{(i)}}|x_{i j^{(i)}},\lambda_i)\right] \nonumber\qquad\qquad\qquad\enspace\enspace\\
    \qquad\qquad\qquad\qquad\qquad\,\times\, p(b_1,\dots,b_n|y,\lambda_1,\dots,\lambda_n),\label{eq:nprobdistri}
\end{gather}
where $i$ labels the branch. On branch $i$, the quantities $a_{i j^{(i)}}\in\{-1,1\}$ and $x_{i j^{(i)}}\in\{1,\dots,k\}$ denote, respectively, the output and the measurement setting of the Alice $A_{i j^{(i)}}$ selected on that branch (for a fixed choice of indices $j^{(i)}$). Bob’s measurement input is $y\in\{1,\dots,k\}$, and his outputs on the $n$ branched subsystems are $b_i\in\{-1,1\}$. The variable $\lambda_i$ is the local hidden variable associated with the independent source connecting Bob and the $i$-th chain of Alices.
\vspace{0.1em}

In general, an $n$-partite correlator in this LHV model is defined as
\begin{gather}
    \langle A_{1 j^{( 1 )}}^{x_{1 j^{( 1 )}}}\dots A_{n j^{( n )}}^{x_{n j^{( n )}}} B^{y} \rangle  =\sum_{\substack{a_{1j^{(1)}},\dots,a_{nj^{(n)}}\\b_1,\dots,b_n}}\left(\prod_{i=1}^{n}a_{i j^{(i)}}\,b_i\right)\nonumber\qquad\\
    \qquad\times P(a_{1 j^{(1)}},\dots,a_{n j^{(n)}},b_1,\dots,b_n|x_{1 j^{(1)}},\dots,x_{n j^{(n)}},y)\nonumber\\
    =\int \prod_{i=1}^n \left[  \mathrm{d}\lambda_i q_i(\lambda_i)\,\sum_{a_{i j^{(i)}}}a_{i j^{(i)}}p(a_{i j^{(i)}}|x_{i j^{(i)}},\lambda_i)\right]\nonumber\,\\
    \qquad\enspace\times\left[\sum_{b_{1},\dots,b_{n}}\left(\prod_{i=1}^n b_i\right) p(b_1,\dots,b_n|y,\lambda_1,\dots,\lambda_n)\right]\nonumber\\
    =\int \left[ \prod_{i=1}^n \mathrm{d}\lambda_i q_i(\lambda_i)\,\langle A_{i j^{( i )}}^{x_{i j^{( i )}}} \rangle_{\lambda_i}\right] \langle B^y \rangle_{\lambda_1,\dots,\lambda_n}\enspace\enspace\quad
    \label{eq:npartcorrelator}
\end{gather}
where 
\begin{align*}
    \langle A_{i j^{(i)}}^{x_{i j^{(i)}}} \rangle_{\lambda_i} &= \sum_{a_{i j^{(i)}}} a_{i j^{(i)}} p(a_{i j^{(i)}}|x_{i j^{(i)}},\lambda_i) \quad\text{and} \\
     \quad \langle B^y \rangle_{\lambda_1,\dots,\lambda_n} &= \sum_{b_{\bar{n}}} b_1 b_2 \dots b_n\, p(b_1,\dots,b_n|y,\lambda_1,\dots,\lambda_n).
\end{align*}
The superscripts indicate measurement settings, while the subscripts specify the position of operators in the network, i.e., the branch index $i$ and the position along the branch, $j^{(i)}$.

\subsection{Extension of Bell Inequalities to The Generalized Star Network \texorpdfstring{$n$}{n}-Locality Scenario}\label{subsec:genstarnlocality}

We now describe a construction that extends a broad family of bipartite Bell inequalities to $n$-local inequalities tailored to the $(n,m,k)$-star network and show the classical bound of the resulting network inequality coincides with that of the original bipartite inequality.

Consider bipartite Bell inequalities of the form
\begin{equation}\label{eq:generalineq}
    I=\sum_{s,\,t=1}^k M_{s t}\,A^s\, B^t\leqslant C
\end{equation}
where $M\in\mathbb{R}^{k\times k}$ is a fixed real matrix. We refer to $M$ as the structure matrix, since it fully specifies the linear combination of correlators $A^s B^t$ appearing in the inequality. The integer $k$ is simultaneously the dimension of $M$ and the number of measurement settings available to each party. For $s,t\in\{1,\dots,k\}$, the observables $A^s$ and $B^t$ denote Alice’s and Bob’s dichotomic measurements, whose outcomes are $a_s\in\{-1,1\}$ and $b_t\in\{-1,1\}$, respectively. The constant $C\geqslant 0$ is the classical upper bound associated with the chosen structure matrix $M$, and it does not depend on the specific outcomes $a_s$ and $b_t$. This compact matrix form covers most standard bipartite Bell inequalities, including the CHSH inequality.

Based on this formalism, the CHSH inequality can be rewritten as
\begin{equation}\label{eq:chshmmatrix}
    M_{\mathrm{CHSH}}=
    \begin{pmatrix}
        1 & 1 \\
        1 & -1
    \end{pmatrix},
    \kern-0.1em 
    \nonumber
\end{equation}

\begin{equation}\label{eq:CHSHmatrixformalism}
    (A^1,\,A^2)\begin{pmatrix}
        1 & 1 \\
        1 & -1
    \end{pmatrix} \begin{pmatrix}
        B^1\\
        B^2
    \end{pmatrix}\leqslant 2.
\end{equation}
For the generalized CHSH inequality~\cite{wehnerTsirelsonBoundsGeneralized2006}, which is used to analyze the $(n,m,k)$-star network nonlocality sharing in~\cite{wangNetworkNonlocalitySharing2022}, takes the form
\begin{equation}\label{eq:genCHSHmatrix}
    \mathbf{A}^T M^{k}_{gCHSH}\mathbf{B}=\mathbf{A}^T\left(J_{1,k}^T-E_{1,k}\right) \mathbf{B}\leqslant2(k-1)
\end{equation}
where $\mathbf{A}^T=(A^1,\dots,A^k)$, $\mathbf{B}=(B^1,\dots,B^k)^T$, the $J_{1,k}^T$ is a $k\times k$ transposed Jordan block with eigenvalue $1$, and $E_{1,k}$ is a $k\times k$ matrix with its $(1,k)$-entry being $1$ and else elements being $0$. For a nontrivial example
\begin{equation}
    M_{gCHSH}^{k=4}=\begin{pmatrix}
        1 & 0 & 0 &-1 \\
        1 & 1 & 0 & 0 \\
        0 & 1 & 1 &0 \\
        0 & 0 & 1 &1
    \end{pmatrix}.\nonumber
\end{equation}

We now state the corresponding $(n,m,k)$-star-network $n$-local inequality:
\begin{gather}\label{eq:verstarnetineq}
    S_{j^{(1)}\dots \,j^{(n)}}^{(n,m,k)}=\sum_{p=1}^k \left| \mathcal{I}^{p}_{j^{(1)}\dots \,j^{(n)}} \right|^{\frac{1}{n}}\leqslant C,
\end{gather}
where $S_{j^{(1)}\dots \,j^{(n)}}^{(n,m,k)}$ is the $(n,m,k)$-star network correlation, and 
\begin{gather}
    \mathcal{I}^p_{j^{(1)}\dots \,j^{(n)}}=\qquad\qquad\qquad\qquad\qquad\qquad\qquad\qquad \nonumber\\ \sum_{x_{1j^{(1)}},\dots,x_{nj^{(n)}}=1}^k \left[\prod_{i=1}^n M_{x_{ij^{( i )}}\,p}\right] \langle A_{1 j^{( 1 )}}^{x_{1 j^{( 1 )}}}\dots A_{n j^{( n )}}^{x_{n j^{( n )}}} B^{p} \rangle.\label{eq:verineqlinegeneralcorrelator}
\end{gather}
Here, $M_{x_{ij^{( i )}}\,p}$ denotes the entry of the structure matrix $M$ in row $x_{ij^{( i )}}$ and column $p$. For each branch $i$, the index $j^{(i)}\in\{1,\dots,m\}$ specifies which Alice $A_{i j^{(i)}}$ is included in the correlator. From a global perspective, the picking function $j:\{1,\dots,n\}\to\{1,\dots,m\},\quad i\mapsto j^{(i)},$ selects exactly one Alice on each branch $i$, so that the set $\{A_{1 j^{(1)}},\dots,A_{n j^{(n)}}\}$ contains precisely one Alice per branch across all $n$ branches.

When the picking function $j$ is fixed, $S_{j^{(1)}\dots j^{(n)}}^{(n,m,k)}$ in~\eqref{eq:verstarnetineq} reduces to the network correlation of a standard star network with $n$ picked Alices—one per branch—and the central Bob. Thus, following theorem 3.1 of~\cite{tavakoliCorrelationsStarNetworks2017a}, the classical $n$-local bound of~\eqref{eq:verstarnetineq} is the same constant $C$ as in the original bipartite inequality~\eqref{eq:generalineq}.

As a concrete example, the $n$-local inequality in~\cite{leeDeviceIndependentInformationProcessing2018}
can be obtained by substituting the structure matrix from~\eqref{eq:genCHSHmatrix} into~\eqref{eq:verineqlinegeneralcorrelator}. After some simplification
\begin{equation}
    \begin{gathered}
        {S_g}_{j^{(1)}\dots \,j^{(n)}}^{(n,m,k)}=\sum_{p=1}^k\left| \mathcal{I}^p_{j^{(1)}\dots \,j^{(n)}} \right|^{\frac{1}{n}}\leqslant 2(k-1)
    \end{gathered}\label{eq:SgCHSHineq}
\end{equation}
in which
\begin{align*}
    \mathcal{I}^p_{j^{(1)}\dots \,j^{(n)}}=\sum_{x_{1j^{(1)}},\dots,x_{nj^{(n)}}=p}^{p+1}\langle A_{1 j^{( 1 )}}^{x_{1 j^{( 1 )}}}\dots A_{n j^{( n )}}^{x_{n j^{( n )}}} B^{p} \rangle,
\end{align*}
and $A_{i j^{(i)}}^{(k+1)}=-A_{i j^{(i)}}^0$.

\section{Quantum Values for the network correlations}
\label{sec:quantumviol}

For the single round of measurements performed by Bob, his joint measurement on all branches can be written in the quantum description as the tensor product $B^y=B_1^y\otimes \dots \otimes B^y_n$. Because each branch is fed by an independent source and Bob measures independently on each subsystem, the resulting multipartite correlator between Alices, one per branch, and Bob factorizes as

\begin{equation}\label{eq:overallcorrelatordecompose}
\langle A_{1 j^{( 1 )}}^{x_{1 j^{( 1 )}}}\dots A_{n j^{( n )}}^{x_{n j^{( n )}}} B^{y} \rangle=\prod_{i=1}^n\langle A_{i j^{( i )}}^{x_{i j^{( i )}}} B_i^y \rangle.
\end{equation}
The bipartite correlator on the $i$-th branch can be calculated as follows:
\begin{gather}
    \langle A_{i j^{( i )}}^{x_{i j^{( i )}}} B_i^y \rangle=\frac{1}{k^{(j^{(i)}-1)}} \sum_{\substack{a_{i1},\dots,a_{ij^{(i)}},b_i,\\
    x_{i1},\dots,x_{i(j^{(i)}-1)} } } a_{i j^{( i )}} b_i\,\times\qquad\qquad\enspace\nonumber \\
    \qquad\qquad\enspace P(a_{i1},\dots,a_{ij^{(i)}},b_i|\,x_{i1},\dots,x_{i(j^{(i)}-1)},y_i).\label{eq:twobodycorrelator}
\end{gather}

The summation runs over all possible outputs of the Alices and Bob, $a_{ip}, b_i \in \{-1,1\}$, and over all measurement settings of the intermediate Alices on branch $i$, $x_{iq} \in \{1,\dots,k\}$ for $q=1,\dots,j^{(i)}-1$. The normalization factor $k^{(j^{(i)}-1)}$ accounts for the uniform choice of the intermediate settings. Thus, the correlator between Bob and the $j^{(i)}$-th Alice on branch $i$ explicitly depends on the entire sequence of preceding weak measurements along that branch.

Initially, on each branch $i$, Bob and the first Alice $A_{i1}$ share the same bipartite state $\rho_{\iota}$, as shown in Fig.~\ref{fig:star_network}. After Bob performs his projective measurement associated with outcome $b_i$ and input $y_i$, the unnormalized post-measurement state reads
\begin{equation}\label{eq:rhoi0withBi}
    \rho_{A_i B_i}^{b_i}=\left(\mathbb{I}\otimes M_{b_i|y_i} \right)\rho_{\iota}\left(\mathbb{I}\otimes M_{b_i|y_i} \right)^{\dagger}
\end{equation}
where $M_{b_i|y_i}$ is a positive operator-valued measurement (POVM) of the form $M_{b_i|y_i}=(\mathbb{I}+b_i \,B_i^y)/2$. Similarly, for Alice we write $M_{a_{ij^{(i)}}|x_{ij^{(i)}}}=(\mathbb{I}+a_{ij^{(i)}} \,A_{ij^{(i)}}^{x_{ij^{(i)}}})/2$. The dichotomic observables of Alices and Bob are parametrized as $A_{ij^{(i)}}^{x_{ij^{(i)}}}={\mathbf{v}_{i j^{( i )}}^{x_{i j^{( i )}}}}^{T}\cdot \boldsymbol{\sigma}$ and
$B_i^y = {\mathbf{w}_{i}^{y}}^{T}\cdot \boldsymbol{\sigma}$, respectively, where ${\mathbf{v}_{i j^{( i )}}^{x_{i j^{( i )}}}}^T$ is the transposed measurement vector for the $j^{(i)}$-th Alice on the $i$-th branch with setting $x_{i j^{(i)}}\in \{1,\dots,k \}$, ${\mathbf{w}_{i}^{y}}^T$ is the transposed measurement vector for Bob on branch $i$ with setting $y\in \{ 1,\dots,k \}$, and $\boldsymbol{\sigma}\coloneq (\sigma_x, \sigma_y, \sigma_z)^T$ is the Pauli vector. By convention, all bold symbols denote column vectors. The symbol $\mathbf{v}$ is reserved for Alices' measurement vectors, while $\mathbf{w}$ is reserved for Bob's measurement vectors.

On Alice's side, tracing out Bob's subsystem yields the unnormalized state
\begin{equation}        
\rho_{A_i}^{bi}=\mathrm{tr}_{B_i}\left( \rho_{A_i B_i}^{b_i} \right)
\end{equation}
which is then passed sequentially along the chain of Alices on branch $i$.

According to Ref.~\cite{silvaMultipleObserversCan2015}, the post-measurement state after a weak measurement by Alice $A_{ij^{(i)}}$ can be written as
\begin{gather}
    \rho_{i j^{(i)}} = \frac{1}{2}F_{i j^{(i)}}\rho_{i(j^{(i)}-1)}+\qquad\qquad\qquad\qquad\qquad\qquad\nonumber\\    
   \frac{1+a_{i j^{(i)}} G_{i j^{(i)}}-F_{i j^{(i)}}}{2}\left[M_{1 \vert x_{i j^{(i)}}} \, \rho_{i ( j^{(i)}-1 )} \, M_{1 \vert x_{i j^{(i)}}} ^\dagger\right]+\nonumber\\
    \frac{1-a_{i j^{(i)}} G_{i j^{(i)}}-F_{i j^{(i)}}}{2}\left[M_{-1 \vert x_{i j^{(i)}}} \, \rho_{i ( j^{(i)}-1 )} \, M_{-1 \vert x_{i j^{(i)}}} ^\dagger\right].\label{eq:optweakmeasuredstate}
\end{gather}
Here $G_{i j^{(i)}}\in[0,1]$ is the precision factor and $F_{i j^{(i)}}\in[0,1]$ is the quality factor of the weak measurement. We call the weak measurement optimal if the two factors satisfy ${G_{i j^{(i)}}}^2+{F_{i j^{(i)}}}^2=1$. The first term represents the undisturbed contribution scaled by the quality factor $F_{i j^{(i)}}$, while the second and third terms describe the contributions associated with outcomes $a_{ij^{(i)}}=+1$ and $a_{ij^{(i)}}=-1$, weighted by the corresponding precision-dependent coefficients. This defines a descending recursive relation along the branch, starting at $j^{(i)}$ and terminating at $\rho_{i0}=\rho_{A_i}^{bi}$. All parties $A_{i1}$ to $A_{i(m-1)}$ perform such optimal weak measurements, whereas the last Alice $A_{im}$ performs a final projective measurement given by
\begin{gather}
    \rho_{i m}= M_{a_{i m} \vert x_{i m}} \, \rho_{i ( m-1 )} \, M_{a_{i m} \vert x_{i m}} ^\dagger .
\end{gather}

We keep all intermediate states unnormalized in order to compute the joint probabilities appearing in Eq.~\eqref{eq:twobodycorrelator}. Specifically, the joint distribution for the outcomes on branch $i$ is obtained by tracing the corresponding unnormalized state:
\begin{equation}\label{eq:pdtraceequality}      P(a_{i1},\dots,a_{ij^{(i)}},b_i|\,x_{i1},\dots,x_{i(j^{(i)}-1)},y_i)=\mathrm{tr}[  \rho_{ij^{(i)}}].
\end{equation}

However, directly evaluating these traces for general sequences of weak measurements is not straightforward. Our main result in this section is the following theorem, which gives a compact analytic expression for the bipartite correlator on each branch and effectively summarizes the cumulative effect of all intermediate measurements.

\begin{theorem*}
    Consider the $i$-th branch in the optimal weak sharing $(n,m,k)$-star network scenario. The observers on this branch are $(A_{im},\,A_{i(m-1)},\,\dots,A_{i1},\, B_i)$, where $B_i$ denotes Bob's measurement on the $i$-th subsystem. Suppose that the initial shared state between $A_{i1}$ and $B$ is the singlet state $\ket{\psi^-}$. Then the bipartite correlator between $A_{ij^{(i)}}$ and $B_i$ is given by
\end{theorem*} 
\begin{align}\label{eq:gencorrelator}
    \langle{A}_{ij^{( i )}}^{x_{i j^{( i )}}} B_{i}^{y} \rangle=-\mathbf{v}_{i j^{( i )}}^{x_{i j^{( i )}} T} \left[\prod_{q=j^{( i )}}^{1} K_{i q} \right] \mathbf{w}_{i}^{y}\,.
\end{align}
where 
\begin{equation}
    K_{i q} =\left\{
    \begin{array}{l l l}
        \enspace\qquad\quad \mathbb{I} &,\, q=j^{(i)}=m \\[0.5ex]
        \,\qquad G_{i j^{(i)}}\,\mathbb{I} &,\, q =j^{(i)}<m \\[0.8ex]
         F_{iq}\,\mathbb{I}+\dfrac{1-F_{iq}}{k}\, T_{iq}&,  \, q<j^{(i)}\quad\quad.
    \end{array} \right.\label{eq:Kiqexpression}
\end{equation}

In Eq.~\eqref{eq:gencorrelator}, the ordered product of matrices is taken from $q=j^{(i)}$ down to $q=1$. This product collects the contributions of all intermediate Alices on branch $i$ between Bob and the chosen Alice $A_{ij^{(i)}}$. In Eq.~\eqref{eq:Kiqexpression}, $\mathbb{I}$ denotes the $3\times3$ identity matrix, and $T_{iq}$ is the measurement matrix associated with the $q$-th Alice on the $i$-th branch, defined as
\begin{equation}\label{eq:measurementmatrix}
    T_{iq}=\sum_{l=1}^k \mathbf{v}_{i q }^{\,l}\mathbf{v}_{i q }^{\,l\enspace T},
\end{equation}
i.e., the sum of all outer products of the $k$ measurement vectors corresponding to the measurement settings of party $A_{iq}$. The proof of the theorem is given in the Appendix.

The three cases in Eq.~\eqref{eq:Kiqexpression} distinguish between (i) the final sharp measurement at $q=m$, (ii) the weak measurement performed by the currently probed Alice $A_{ij^{(i)}}$ when $j^{(i)}<m$, and (iii) the weak measurements performed by all earlier Alices $A_{iq}$ with $q<j^{(i)}$.

The expression~\eqref{eq:gencorrelator} is general for the optimal weak-sharing scenario considered here: it holds for arbitrary choices of measurement settings for Alices and Bob, as long as the initial state on each branch is the singlet. The entire dependence on the weak-measurement strengths and on the local measurement settings is encoded in the matrices $K_{iq}$ and $T_{iq}$.

Within this framework, we can recover as a special case the $(n,m,k)$-star-network scenario studied in Ref.~\cite{wangNetworkNonlocalitySharing2022}. There, the Alices' and Bob's measurement vectors are chosen as 
\begin{gather*}
    {\mathbf{v}_{ij^{(i)}}^{x_{i j^{(i)}}}}^T=\left( \sin\frac{(x_{i j^{(i)}}-1)\pi}{k},0, \cos\frac{(x_{i j^{(i)}}-1)\pi}{k} \right)
\end{gather*}
and 
\begin{gather*}
    {\mathbf{w}_i^y}^T=\left( \sin\frac{(2y-1)\pi}{2k},0, \cos\frac{(2y-1)\pi}{2k} \right),
\end{gather*}
which, when inserted into our general formula, reproduces the bipartite correlator $(\mathrm{A}6)$ of Ref.~\cite{wangNetworkNonlocalitySharing2022}. Here, for consistency with our present notation, we use $x_{i j^{(i)}},\,y\in \{1,\dots, k \}$, whereas in Ref.~\cite{wangNetworkNonlocalitySharing2022} the inputs range from $0$ to $k-1$; the measurement vectors are therefore shifted accordingly. For these specific settings, the intermediate measurement matrices $T_{iq}$ are independent of the branch index $i$ and the position $q$, and take the simple form

\begin{equation}
    T_{iq}=\begin{pmatrix}
        \frac{1}{2}k & 0 &0\\
        0 & 0 & 0\\
        0 & 0 & \frac{1}{2}k
    \end{pmatrix}.
\end{equation}
Substituting these matrices into Eq.~\eqref{eq:gencorrelator} and simplifying the resulting trigonometric expressions, one obtains
\begin{equation}
      \langle A_{i j^{( i )}}^{x_{i j^{( i )}}} B_{i}^{y} \rangle=\left( -\cos\left[  \frac{\pi(2y-2 x_{i j^{(i)}}+1)}{2k}  \right] \right) \prod_{q=1}^{j^{(i)}}T_{iq}
\end{equation}
where
\[
T_{i q}=\left\{
    \begin{array}{l l l}
        \enspace\quad 1 &,\, q=j^{(i)}=m \\[0.5ex]
        \quad G_{i q} &,\, q =j^{(i)}<m \\[0.8ex]
        \enspace \frac{1+F_{iq}}{2} &,  \, q<j^{(i)}\quad\quad.
    \end{array} \right.
\]
Finally, inserting the above bipartite correlator into the corresponding network expression~\eqref{eq:SgCHSHineq}, we obtain the quantum value of the generalized CHSH-type star-network correlation:

\begin{equation}
    \begin{gathered}
        {S_g}_{j^{(1)}\dots \,j^{(n)}}^{(n,m,k)}=k\cos\left( \frac{\pi}{2k}\right) \left( \prod_{i=1}^n \prod_{q=1}^{j^{(i)}} T_{iq}\right)^{\frac{1}{n}}.
    \end{gathered}
\end{equation}
This reproduces the quantum correlation for $n$-locality given in Eq.~(11) of Ref.~\cite{wangNetworkNonlocalitySharing2022}.

\section{Network nonlocality sharing Framework and Case for Vertesi's inequalities}\label{sec:EGvertesi}

\subsection{Network Nonlocality Sharing Framework for Bipartite Bell inequalities}
We introduce a framework for network nonlocality sharing constructed from a given bipartite Bell inequality of the form~\eqref{eq:generalineq} in the \((n,m,k)\)-star network. The $(n,m,k)$-star network nonlocality sharing can be achieved when the network inequality~\eqref{eq:verstarnetineq} is violated by the quantum values of the network correlations for the parties \(\left(A_{1 j^{(1)}},\dots,A_{n j^{(n)}},B\right)\), for every admissible choice of \(j\). Concretely, for network nonlocality sharing in the \((2,2,k)\) case, the quantum network correlations corresponding to the four groups \((A_{11},A_{21},B)\), \((A_{11},A_{22},B)\), \((A_{12},A_{21},B)\), and \((A_{12},A_{22},B)\) should all simultaneously violate the corresponding network Bell inequality.

To determine the quantum values of the relevant star-network correlations, we proceed as follows. First, we evaluate the bipartite correlators using Eq.\eqref{eq:gencorrelator}, choosing measurement settings (or measurement vectors) for the Alices and Bob—for example, those that maximize the bipartite quantum violation. Next, substituting these correlators into Eq.~\eqref{eq:overallcorrelatordecompose} yields the multipartite correlators. Finally, the star-network correlation (the left-hand side of inequality~\eqref{eq:verstarnetineq}) is computed by inserting the resulting multipartite correlators and the structure matrix of the initially specified bipartite Bell inequality into Eq.~\eqref{eq:verineqlinegeneralcorrelator}.

\subsection{Network Nonlocality Sharing for Vertesi's Inequalities}
We now follow the steps to study the $(n,m,k)$-star network nonlocality sharing from Vértesi inequalities~\cite{vertesiMoreEfficientBell2008}. Vértesi’s construction directly addresses an open question posed by Gisin~\cite{gisinBellInequalitiesMany2007} in 2007, by demonstrating that there exist Bell inequalities that are strictly more efficient than the CHSH inequality for Werner states, in the sense that the ratio of quantum value over classical value of corresponding correlation is more than $\sqrt{2}$. For convenience, the ratio will be referred to as $Q/C$ in the rest of the context. Although these inequalities require a comparatively large number of measurement settings and are therefore less convenient for experimental implementation, they play an important conceptual role in clarifying the boundary between local and nonlocal correlations. 
\vspace{-0.4em}

\subsubsection{The Vértesi Generalized Star Network \texorpdfstring{$n$}{n}-Locality}

In~\cite{vertesiMoreEfficientBell2008}, Vértesi introduces a family of Bell correlators for the bipartite scenario,

\begin{gather}
    I_{n_A,n_B}=\sum_{s=1}^{n_A} \sum_{t=1}^{n_B}A^s B^t + \sum_{1\leqslant s<t \leqslant n_B}A^{st}(B^s -B^t)\nonumber\\
    +\sum_{1\leqslant s<t \leqslant n_A}B^{st}(A^s -A^t).\label{eq:vertesieq}
\end{gather}
The total number of measurement settings for Alice is $m_A$, consisting of $n_A$ observables $A^s$ and $n_B(n_B-1)/2$ observables $A^{st}$. For clarity, the $A^{st}$ are also distinct measurement settings, whose labels $(s,t)$ are ordered lexicographically. Similarly, Bob has $m_B = n_B + n_A(n_A-1)/2$ measurement settings. In the following, we will mainly focus on the symmetric case $n_A = n_B = \tilde{n}$; we refer to $\tilde{n}$ as the Vértesi parameter. In this symmetric case, the classical upper bound of~\eqref{eq:vertesieq} is $\tilde{n}^2$, i.e.,
\begin{equation}
I_{\tilde{n},\tilde{n}}\leqslant\tilde{n}^2.
\end{equation}
The quantity $I_{\tilde{n},\tilde{n}}$ can be rewritten in the form of~\eqref{eq:generalineq}, with the structure matrix
\begin{equation}
    M_V = \left(\begin{array}{@{}c|c@{}}
  \mathbf{1}
  & N \\
\hline
  N^T & \mathbf{0}
\end{array}\right),\label{eq:Mv1}
\end{equation}
where $\mathbf{1}$ is the $\tilde{n}\times\tilde{n}$ all-ones matrix and $\mathbf{0}$ is the $\dfrac{1}{2}\tilde{n}(\tilde{n}-1)\times\dfrac{1}{2}\tilde{n}(\tilde{n}-1)$ zero matrix. To specify the entries of $N$, which is a $\tilde{n}\times\dfrac{1}{2}\tilde{n}(\tilde{n}-1)$ matrix, we need an explicit correspondence between the lexicographic labels $st$ in~\eqref{eq:vertesieq} and the row and column indices of $N$. In our setup, Alice’s measurement settings are ordered as $A^1,\dots,A^{\tilde{n}}$, followed by $A^{\tilde{n}+1}=A^{12}$ and, more generally, $A^{\tilde{n}+r}=A^{1\,(r+1)}$ for all $r\leqslant\tilde{n}-1$. These are then followed by $A^{\tilde{n}+\tilde{n}}=A^{23}$, etc. Following this iterative rule, one can verify that for every sequential label $l\in\{\tilde{n}+1,\dots,\tilde{n}+\tilde{n}(\tilde{n}-1)/2\}$ there exists a unique ordered pair $(s,t)$ with $1\leqslant s<t\leqslant\tilde{n}$ such that $l=s[\tilde{n}-(s+1)/2]+t$. Based on this one-to-one correspondence between the two labeling schemes, the entries of $N$ are defined by
\begin{gather}
    N_{j\,(l-\tilde{n})} = 
\begin{cases} 
      \enspace1, & \text{if } j = s_l \\
     -1, & \text{if } j = t_l \\
      \enspace0, & \text{else}
\end{cases}
\end{gather}
where $s_l$ and $t_l$ form the unique pair $(s_l,\,t_l)$ such that $l=s_l [\tilde{n}-(s_l+1)/2]+t_l$.

To make the construction more transparent, we now present two explicit examples. For $\tilde{n}=2$, the total number of settings is $k=3$, and the structure matrix reads
\begin{equation}
    M_V^{k=3}=\left(\begin{array}{@{}cc|c@{}}
  1 & 1  &1\\
  1 & 1 & -1\\
\hline
  1 &-1 &0
\end{array}\right)\label{eq:mvk=3}.
\end{equation}
For $\tilde{n}=3$ (so that $k=6$), one obtains
\begin{equation}
    M_V^{k=6}=\left(\begin{array}{@{}ccc|ccc@{}}
        1 & 1 & 1 & 1& 1& 0\\
        1& 1 &1 &-1 &0 &1\\
        1 & 1 & 1 &0 &-1 &-1\\
        \hline
        1 & -1&0 & 0& 0&0\\
        1 & 0&-1 & 0& 0&0\\
        0 & 1&-1 & 0& 0&0
    \end{array}\right),
\end{equation}
where the star-network generalizations corresponding to these two cases will be analyzed analytically in Sec.~\ref{sec:networknonlocalityanalysis}.

We now turn to the extension from bipartite locality to $n$-locality for the Vértesi inequality in the star-network configuration. Since the symmetric Vértesi inequality can indeed be cast in the form~\eqref{eq:generalineq}, it follows from the discussion in Sec.~\ref{subsec:genstarnlocality} that

\begin{gather}
S_{j^{(1)}\dots \,j^{(n)}}^{(n,m,k)}=\sum_{p=1}^k \left| \mathcal{I}^p_{j^{(1)}\dots \,j^{(n)}} \right|^{\frac{1}{n}}\leqslant \tilde{n}^2,
\end{gather}
where $k=\tilde{n}(\tilde{n}+1)/2$, and the specific Vértesi structure matrix $M_V$ is substituted into the expression~\eqref{eq:verineqlinegeneralcorrelator} defining $\mathcal{I}^p_{j^{(1)}\dots \,j^{(n)}}$.

\subsubsection{Network Nonlocality Sharing}\label{sec:networknonlocalityanalysis}

To evaluate the final correlations, we first specify the measurement settings. In contrast to the generalized CHSH scenario considered in~\cite{wangNetworkNonlocalitySharing2022}, where analytical expressions for the quantum correlations are available, only a few values of $\tilde{n}$ in the Vértesi star-network setting can be examined analytically. This limitation arises from the difficulty of obtaining the exact maximal quantum violation of the corresponding Vértesi inequality. 

According to Vértesi's analysis~\cite{vertesiMoreEfficientBell2008}, the correlator~\eqref{eq:vertesieq} on the singlet state $\rho=\ket{\psi^-}\bra{\psi^-}$ can be written as
\begin{equation}
    I^d=\left| \sum_{s,t=1}^{k} {M_V}_{st}\,{\mathbf{v}}^s\cdot {\mathbf{w}}^t \right|,\label{eq:Id1}
\end{equation}
where $M_V$ is the structure matrix in the form~\eqref{eq:Mv1}, and $d$ denotes the dimension of the unit vectors ${\mathbf{v}}^s$ and ${\mathbf{w}}^t$. In our setting, ${\mathbf{v}}^s$ and ${\mathbf{w}}^t$ are the measurement vectors for Alice and Bob, respectively, i.e., $d=3$. Vértesi showed that the maximal value of~\eqref{eq:Id1} is equivalent to the optimization of
\begin{equation}
    \max\left(  \left| \sum_{s=1}^{\tilde{n}} {\mathbf{v}}^s \right|^2 +2 \sum_{1\leqslant s<t \leqslant \tilde{n}} |{\mathbf{v}}^s-{\mathbf{v}}^t|  \right)\label{eq:maxId1}.
\end{equation}
This equivalent expression is attained by choosing Alice’s and Bob’s first $\tilde{n}$ measurement vectors to coincide, $\mathbf{v}^s=\mathbf{w}^s$, and by setting the remaining $\tilde{n}(\tilde{n}-1)/2$ measurement directions as ${\mathbf{w}}^{st}={\mathbf{v}}^{st}={\mathbf{v}}^s-{\mathbf{v}}^t$, where the index $st$ follows the lexicographic order. For $\tilde{n}=2$ and $\tilde{n}=3$, the expression~\eqref{eq:maxId1} attains its maximum values $5$ and $12$, respectively. In these cases, the corresponding configuration of measurement vectors only requires that the pairwise angles between the $\tilde{n}$ primary vectors are all equal to $\pi/6$. When $\tilde{n}=4$, however, such a configuration becomes impossible in three dimensions, as no fourth unit vector can simultaneously maintain an angle of $\pi/6$ with each of the others. Consequently, one must resort to numerical methods to approximate an optimal set of vectors. In the present work, we employ exactly the same vector configurations as those introduced in~\cite{vertesiMoreEfficientBell2008}.

For simplicity, we mainly focus on star networks with two branches and two Alices per branch, i.e., $n=2$ and $m=2$. Furthermore, we assume that all Alices perform identical optimal weak measurements, so that $G_{ij^{(i)}}=G$ and $F_{ij^{(i)}}=F$ for all $i,j^{(i)}$. On both branches, Bob implements the same measurement settings, $B^y_i=B^y$, and the Alices’ observables satisfy ${A}_{ij^{( i )}}^{x_{i j^{( i )}}}={A}_{j}^{x_{j}}$.

We first analyze the case $\tilde{n}=2$ (corresponding to $k=3$). To simplify the calculations, we choose the following measurement settings, which are known to yield the maximal bipartite quantum value $5$:
\begin{align}
    {\mathbf{v}^1}^{\,T} =(1,0,0),\,{\mathbf{v}^2}^{\,T} =(\frac{1}{2},\frac{\sqrt{3}}{2},0),\,{\mathbf{v}^{12}}^{\,T} =(\frac{1}{2},-\frac{\sqrt{3}}{2},0)\nonumber
\end{align}
and $\mathbf{w}^s = \mathbf{v}^s$ for $s\in\{1,2,12\}$, following the lexicographic order.
\begin{figure*}[t] 
    \centering
    \includegraphics[width=\textwidth]{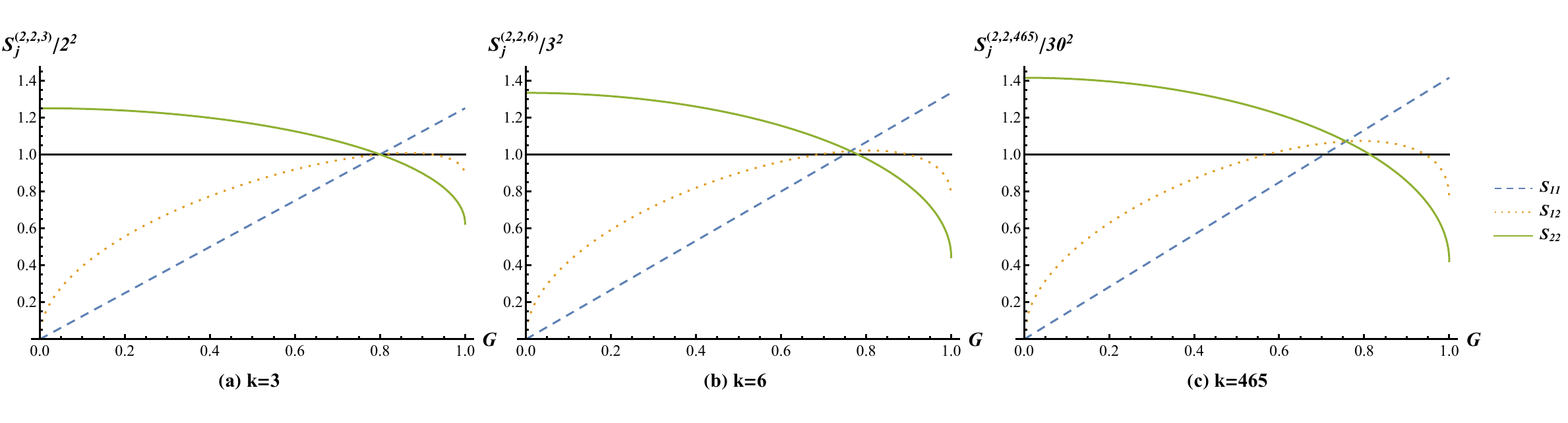}
    \vspace{-3em}
   \caption{Normalized quantum correlations $S_j$ in the $(2,2,k)$-star network as a function of the precision factor $G$ for varying number of measurement settings $k$. The plots illustrate three correlations—$S_{11}$ (dashed blue line) for Alice$_{11}$ and Alice$_{21}$, $S_{12}$ (dotted gold line) for Alice$_{11}$ and Alice$_{22}$, and $S_{22}$ (solid green line) for Alice$_{12}$ and Alice$_{22}$—plotted against the precision factor, $G$. The vertical axis represents the correlation values $S_{j}$ normalized by the corresponding classical bound $\tilde{n}^2$, with the classical bound indicated by a solid black reference line at $1.0$. Each panel corresponds to a $(2, 2, k)$-star network scenario with a different number of measurement settings: (a) $k = 3$, (b) $k = 6$, and (c) $k = 465$. The plots demonstrate that a simultaneous violation of the classical bound is possible when $k>3$, and the quantum network nonlocality sharing effect becomes more pronounced as the number of measurement settings $k$ increases.}
    \label{fig:latex_composed_figure}
\end{figure*}


Here we provide a sample calculation for $S_{12}^{(2,2,3)}$. Since all Alices perform identical optimal weak measurements and use the same settings on both branches, it is immaterial whether we select $A_{11}$ on the first branch and $A_{22}$ on the second, or $A_{12}$ and $A_{21}$. In particular, we have $S_{12}^{(2,2,3)}=S_{21}^{(2,2,3)}$. Following the general procedure for evaluating the network nonlocality, we first compute the bipartite correlators in terms of the measurement vectors using the formula~\eqref{eq:gencorrelator}:
\begin{equation}
    \langle A_{1 1}^{x_{1 1}} B^{p}_1\rangle=-G\,\mathbf{v}_{11}^{x_{11} T}  \mathbf{w}_{1}^{p}
\end{equation}
where $K_{11}=G$.
\begin{align}
    \langle A_{2 2}^{x_{2 2}} B^{p}_2 \rangle &=
    -\mathbf{v}_{11}^{x_{11} T} K_{22}K_{21} \mathbf{w}_{1}^{p}
\end{align}
where $K_{22}=1$ and $K_{21}=F\,\mathbb{I}+\dfrac{1-F}{3}\, T_{21}$. Since all Alices are constrained to use the same measurement settings, we have
\begin{gather}
    T_{21}=T=\mathbf{v}^1 {\mathbf{v}^1}^T+\mathbf{v}^2 {\mathbf{v}^2}^T+\mathbf{v}^{12} {\mathbf{v}^{12}}^T\nonumber\\
    =\begin{pmatrix}
        \frac{3}{2} & 0 & 0\\
        0 & \frac{3}{2} &0\\
        0 & 0 & 0
    \end{pmatrix}.
\end{gather}
We then substitute the matrix $M_V^{k=3}$~\eqref{eq:mvk=3}, together with the above bipartite correlators, into~\eqref{eq:verineqlinegeneralcorrelator}:
\begin{align}
    |\mathcal{I}^p_{12}|&=
    \sum_{x_{11},x_{22}=1}^3 \left[\prod_{i=1}^2 [M^{k=3}_V]_{x_{ij^{( i )}}\,p}\right]\langle A_{1 1}^{x_{1 1}} B^{p}_1\rangle \langle A_{2 2}^{x_{2 2}} B^{p}_2 \rangle \label{eq:Ip12analyticexp},
\end{align}
which yields
\begin{align}
    |\mathcal{I}^1_{12}| &= 2G(1+F)=2G\,(1+\sqrt{1-G^2})\nonumber\\
    |\mathcal{I}^2_{12}| &=2G\,(1+\sqrt{1-G^2})\\
    |\mathcal{I}^3_{12}| &=\frac{1}{2}G\,(1+\sqrt{1-G^2}).\nonumber
\end{align}
We write these quantities as functions of $G$ because, in the next step, we will plot $S_{12}^{(2,2,3)}$ as a function of $G$ to investigate whether simultaneous violations occur. Substituting all three values $|\mathcal{I}^p_{12}|$ into the definition of the multipartite correlation~\eqref{eq:verstarnetineq}, we obtain
\begin{align}
    S_{12}^{(2,2,3)}&=\sqrt{|\mathcal{I}^1_{12}|}+\sqrt{|\mathcal{I}^2_{12}|} +\sqrt{|\mathcal{I}^3_{12}|}\nonumber\\
    &= \frac{5\sqrt{2}}{2}\sqrt{G(1+\sqrt{1-G^2})}
\end{align}
Analogously, one can determine the remaining quantities. For $A_{11}$ and $A_{21}$, we find
\begin{gather}
    |\mathcal{I}^1_{11}| = 4G^2,\enspace
    |\mathcal{I}^2_{11}| =4G^2,\enspace
    |\mathcal{I}^3_{11}| =G^2,
\end{gather}
and the corresponding multipartite correlation is
\begin{align}
    S_{11}^{(2,2,3)}=5\,G.
\end{align}
For $A_{12}$ and $A_{22}$,
\begin{align}
    |\mathcal{I}^1_{22}|= &|\mathcal{I}^2_{22}|= (1+\sqrt{1-G^2})^2\nonumber,\\
    |\mathcal{I}^3_{22}| &=\frac{1}{4}(1+\sqrt{1-G^2})^2,
\end{align}
and the corresponding multipartite correlation is
\begin{align}
        S_{22}^{(2,2,3)}&=\frac{5}{2}(1+\sqrt{1-G^2}).
\end{align}
These three functions are plotted in panel (a) of Fig.~\ref{fig:latex_composed_figure}. All curves are normalized by the corresponding classical bound $\tilde{n}^2$, which facilitates a direct comparison of the violation strength. From the figure, we see that although each $S_j$ can exceed the classical bound for some values of $G$, there is no value of $G$ for which all three correlations simultaneously violate the $n$-locality inequality; in other words, no $n$-nonlocality sharing occurs in this case. The point of intersection of the three curves is located at $G=\dfrac{4}{5}$ and $S_j=4$, which coincides with the classical bound.

We next consider the case $\tilde{n}=3$ (so that $k=6$). The measurement settings for Alices are chosen as
\begin{gather}
    {\mathbf{v}^1}^{\,T} =(0,\frac{\sqrt{6}}{3},\frac{\sqrt{3}}{3}),\qquad{\mathbf{v}^2}^{\,T} =(\frac{1}{2},\frac{\sqrt{6}}{3},-\frac{\sqrt{3}}{6}),\nonumber\\{\mathbf{v}^{3}}^{\,T} =(-\frac{1}{2},\frac{\sqrt{6}}{3},-\frac{\sqrt{3}}{6}).\nonumber
\end{gather}
$\mathbf{v}^{st}=\mathbf{v}^s - \mathbf{v}^t$, and for Bob we again take $\mathbf{w}^s = \mathbf{v}^s$ for $s\in\{1,2,3,12,13,23\}$ in lexicographic order. This choice of vectors ensures that the maximal bipartite quantum value of $12$ is achieved.

The resulting network correlations are
\begin{align}
    S_{11}^{(2,2,6)} &= 12\,G\nonumber\\
    S_{12}^{(2,2,6)} &= 4\sqrt{3}\,\sqrt{G(1+2\sqrt{1-G^2})} \nonumber\\
    S_{22}^{(2,2,6)}&=4\, (1+2\sqrt{1-G^2})
\end{align}
and their behavior is shown in panel (b) of Fig.~\ref{fig:latex_composed_figure}. One observes a narrow interval around $G\approx 0.76$ in which all three correlations exceed the classical bound simultaneously; in this parameter region, the network nonlocality sharing is thus achieved.

Finally, we consider the measurement settings introduced in~\cite{vertesiMoreEfficientBell2008} for the smallest value $\tilde{n}=30$ whose bipartite quantum violation ratio (quantum value over classical value of a correlation) exceeds the previously well-known CHSH ratio $\sqrt{2}$. These measurement settings are constructed as follows. One first projects three-dimensional unit vectors onto the $xOy$ plane. Since each measurement vector has unit length, each point in the $xOy$ plane (with radius bounded by one) uniquely specifies a measurement direction. Three concentric circles $\mathcal{C}_I$, $\mathcal{C}_{II}$, and $\mathcal{C}_{III}$ are then chosen with radii $\rho_{\mathrm{I}}=22/100$, $\rho_{\mathrm{II}}=52/100$, and $\rho_{\mathrm{III}}=77/100$, respectively. On $\mathcal{C}_I$, starting from the point $(\rho_I \cos\dfrac{\pi}{4},\rho_I \sin \dfrac{\pi}{4})$, one picks $4$ points by successive rotations of angle $\pi/2$. On $\rho_{II}$, starting from $(0,\rho_{II})$ with rotation angle $\pi/5$, one obtains $10$ points, and on $\mathcal{C}_{III}$, starting from $(0,\rho_{III})$ and rotating by $\pi/8$, one obtains $16$ points. Altogether, these measurement settings yield the ratio $I^3_{\tilde{n},\tilde{n}}/\tilde{n}^2=1.415199$. Using this configuration, we compute and plot $S^{(2,2,465)}_{j^{(1)}j^{(2)}}$ for $\tilde{n}=30$ (i.e., $k=465$), shown in panel (c) of Fig.~\ref{fig:latex_composed_figure}.

In~\cite{wangNetworkNonlocalitySharing2022}, it was found that, within their optimal weak star-network scenario based on generalized CHSH inequalities, the $Q/C$ value decreases as the number of measurement settings $k$ increases. This is because the corresponding ratio takes the form $k \cos(\dfrac{\pi}{2k})/(k-1)$, which is a monotonically decreasing function of $k$. Thus, there is no nonlocality sharing among all observers for $k>3$. In contrast, in the star-network extension of Vértesi inequality considered here, the network nonlocality sharing can still be witnessed for a large number of settings.

%
\section{Conclusion and Discussion}
\label{sec:summary}

We have developed a framework with given bipartite Bell inequalities for studying nonlocality sharing in \((n,m,k)\)-star networks under the source-independence assumption. The bipartite correlator in our framework is expressed explicitly in terms of the local measurement vectors and the weak-measurement parameters, and it is valid for arbitrary values of \(n\), \(m\), and \(k\), provided that each source initially distributes a singlet state. This analytic expression substantially simplifies the computation of network correlations, in particular for scenarios with large numbers of measurement settings.

Within the framework, we applied the construction to the symmetric Vértési inequalities and obtained explicit expressions for the corresponding network quantities in the \((2,2,3)\) and \((2,2,6)\) cases, as well as numerical results for $(2,2,465)$ case. 
Network nonlocality sharing is witnessed in $(2,2,6)$ and $(2,2,465)$ cases, with the latter case having a more robust sharing ability.

The findings in the Vértési case should be regarded as illustrative rather than optimal. We did not perform a full optimization over measurement choices or over families of inequalities, and the Vértési inequalities serve here primarily to demonstrate that the framework remains effective in nontrivial multi-setting scenarios. Moreover, the structure matrix of the Vértési inequalities is relatively sparse, suggesting that bipartite Bell inequalities with denser structure matrices may exploit the available network resources more efficiently and yield more robust violations within the same weak-sharing scheme. Our analysis also adopts several simplifying assumptions, most notably maximally entangled singlet states on each branch and an optimal weak-measurement model with identical parameters for all Alices. Extending the framework to noisy or partially entangled states, and to alternative weak- or mixed-measurement schemes, therefore, constitutes a natural next step.

\begin{acknowledgments}
This work is supported by the National Natural Science Foundation of China (Grants No. 12165020).
\end{acknowledgments}

\onecolumngrid
\vspace{1.5em}

\appendix

\phantomsection
\addcontentsline{toc}{section}{Appendix: Proof of Theorem on the bipartite correlator}
\section*{Appendix: Proof of Theorem on the bipartite correlator}
\label{app:proof}

\setcounter{equation}{0}                 
\renewcommand{\theequation}{A\arabic{equation}} 

We begin with the simplification of $\rho_{A_i}^{bi}$. Since the initial shared state between $A_{i1}$ and $B_i$ is the singlet state, we represent its density matrix as
\begin{gather}\label{eq:singletsigmaexpansion}
    \rho_{\iota}=\rho_{\psi^-}=\frac{1}{4}\left( \mathbb{I}\otimes\mathbb{I}-\sigma_x \otimes \sigma_x-\sigma_y \otimes \sigma_y-\sigma_z \otimes \sigma_z\right).
\end{gather}
Substituting this expression into~\eqref{eq:rhoi0withBi} and taking the partial trace $\mathrm{tr}_{B_i}$, we obtain a simplified form of $\rho_{i0}$:
\begin{equation}\label{eq:rhoi0expanded}
    \rho_{i0}=\rho_{A_i}^{bi}=\frac{1}{4}\left( \mathbb{I}-b_i B^y_i \right).
\end{equation}
Next, we substitute~\eqref{eq:pdtraceequality} into~\eqref{eq:twobodycorrelator} and sum over $a_{i j^{(i)}}$:
\begin{align}
    \sum_{\substack{a_{i j^{(i)}}\in \\\{-1,1\}}} a_{i j^{(i)}} \, \mathrm{t r} ( \rho_{i j^{(i)}} ) &= \sum_{\substack{a_{i j^{(i)}}\in \\ \{-1,1\}}} a_{i j^{(i)}} \mathrm{tr} \left\{ \frac{1}{2}F_{i j^{(i)}}\rho_{i(j^{(i)}-1)}+ 
   \frac{1+a_{i j^{(i)}} G_{i j^{(i)}}-F_{i j^{(i)}}}{2}\left[M_{1 \vert x_{i j^{(i)}}} \, \rho_{i ( j^{(i)}-1 )} \, M_{1 \vert x_{i j^{(i)}}} ^\dagger\right]\right.\nonumber
   \\
   &\quad \quad\quad\quad+\left.\frac{1-a_{i j^{(i)}} G_{i j^{(i)}}-F_{i j^{(i)}}}{2}\left[M_{-1 \vert x_{i j^{(i)}}} \, \rho_{i ( j^{(i)}-1 )} \, M_{-1 \vert x_{i j^{(i)}}} ^\dagger\right] \right\}\nonumber\\
   &=G_{i j^{(i)}}\mathrm{tr}\left[\frac{1}{2}\left( \mathbb{I}+A_{ij^{(i)}}^{x_{ij^{(i)}}} \right) \rho_{i (j^{(i)}-1)} \right] -G_{i j^{(i)}}\mathrm{tr}\left[\frac{1}{2}\left( \mathbb{I}-A_{ij^{(i)}}^{x_{ij^{(i)}}} \right) \rho_{i (j^{(i)}-1)} \right]\nonumber\\
   &=G_{i j^{(i)}}\mathrm{tr}\left[A_{ij^{(i)}}^{x_{ij^{(i)}}}\rho_{i (j^{(i)}-1)} \right]\label{eq:JianhuiA3}
\end{align}
where we have expanded $M_{1 \vert x_{i j^{(i)}}}$ and $M_{-1 \vert x_{i j^{(i)}}}$ in terms of $A_{ij^{(i)}}^{x_{ij^{(i)}}}$ and applied the cyclic property of trace. Furthermore, $\sum_{a_{i j^{(i)}}} f(a_{i j^{(i)}})=0$ if $f$ is an odd function of $a_{i j^{(i)}}$. A similar calculation establishes that $\sum_{a_{i m}} a_{i m}\, \mathrm{tr}\left( \rho_{im} \right)=\mathrm{tr}\left[ A_{im}^{x_{im}} \rho_{i (m-1)} \right]$ for the final node's projective measurements. We now state the following claim.

\begin{claim*}
For an arbitrary vector $\mathbf{u}$ independent of $a_{i j^{(i)}}$ and $x_{i j^{(i)}}$, and an arbitrary state $\rho_{i j^{(i)}}$ after weak measurement with $j^{(i)}< m$,
\end{claim*}
\begin{gather}
    \frac{1}{k}\sum_{x_{i j^{(i)}}=1}^k \sum_{a_{i j^{(i)}}=-1}^1 \mathrm{tr}\left[\mathbf{u}^T \boldsymbol{\sigma} \rho_{i j^{(i)}}  \right]=F_{i j^{(i)}}\,\mathrm{tr}\left[ \mathbf{u}^T\boldsymbol{\sigma}\rho_{i (j^{(i)}-1)} \right]
    +\frac{1}{k}\left[1- F_{i j^{(i)}} \right]\mathrm{tr}\left[ \mathbf{u}^T T_{i j^{(i)}}\boldsymbol{\sigma}\rho_{i (j^{(i)}-1)} \right].\label{eq:mainrecursiveR}
\end{gather}
\begin{proof}
    Substituting the state after weak measurement~\eqref{eq:optweakmeasuredstate} into the left-hand side of~\eqref{eq:mainrecursiveR}, the first summand after the $\dfrac{1}{k}$-summation becomes
\begin{equation}\label{eq:mainrecursiveR1}
    \frac{1}{k}\sum_{a_{i j^{(i)}},\,
    x_{i j^{(i)}}}\mathrm{tr}\left[\mathbf{u}^T \boldsymbol{\sigma} \frac{F_{i j^{(i)}}}{2}\rho_{i (j^{(i)}-1)}  \right]=F_{i j^{(i)}}\,\mathrm{tr}\left[ \mathbf{u}^T\boldsymbol{\sigma}\rho_{i (j^{(i)}-1)} \right].
\end{equation}
The equality holds because there are $2k$ terms independent of $a_{i j^{(i)}}$ and $x_{i j^{(i)}}$. For the second and third summands, we focus on their sum. Defining 
\begin{equation}
    \widetilde{A}_1=M_{1 \vert x_{i j^{(i)}}} ^\dagger \mathbf{u}^T \boldsymbol{\sigma}\, M_{1 \vert x_{i j^{(i)}}}
\qquad\qquad\text{and}\qquad\qquad
    \widetilde{A}_{-1}=M_{-1 \vert x_{i j^{(i)}}} ^\dagger \mathbf{u}^T \boldsymbol{\sigma}\, M_{-1 \vert x_{i j^{(i)}}},
\end{equation}
the sum of the second and third summands can be expressed compactly as
\begin{gather}
    \frac{1}{k}\sum_{a_{i j^{(i)}},\,
    x_{i j^{(i)}}}
    \left\{ 
    \frac{1+a_{i j^{(i)}}G_{i j^{(i)}}-F_{{i j^{(i)}}}}{2}\mathrm{tr}
    \left[
    \widetilde{A}_1 \rho_{i (j^{(i)}-1)}
    \right] 
     \right.
     +\left. \frac{1-a_{i j^{(i)}}G_{i j^{(i)}}-F_{{i j^{(i)}}}}{2}\mathrm{tr}
    \left[
    \widetilde{A}_{-1} \rho_{i (j^{(i)}-1)}
    \right]
     \right\}.
\end{gather}
All $a_{i j^{(i)}}$ terms vanish upon summation over $a_{i j^{(i)}}\in\{-1,1\}$, allowing us to reduce the expression to

\begin{gather}
    \frac{1}{k}\sum_{\substack{a_{i j^{(i)}},\\
    x_{i j^{(i)}}}}
    \left\{ \frac{1-F_{{i j^{(i)}}}}{2}\mathrm{tr}\left[ 
    \left(\widetilde{A}_{1}+\widetilde{A}_{-1} \right)\rho_{i (j^{(i)}-1)} \right] \right\}
    = \frac{1}{k}\sum_{\substack{a_{i j^{(i)}},\\
    x_{i j^{(i)}}}}
    \left\{ \frac{1-F_{{i j^{(i)}}}}{4}\mathrm{tr}\left[ 
    \left(
    \mathbf{u}^T \boldsymbol{\sigma}+A_{ij^{(i)}}^{x_{ij^{(i)}}}\mathbf{u}^T \boldsymbol{\sigma} A_{ij^{(i)}}^{x_{ij^{(i)}}}
    \right)\rho_{i (j^{(i)}-1)} \right] \right\}.\label{eq:coreqmiddlestep1}
\end{gather}
We now evaluate the second summand within the trace in~\eqref{eq:coreqmiddlestep1}. Since $A_{ij^{(i)}}^{x_{ij^{(i)}}}={\mathbf{v}_{i j^{( i )}}^{x_{i j^{( i )}}}}^{T}\cdot \boldsymbol{\sigma}$, we first compute
\begin{gather}\label{eq:sandwitchedPauliequa}
    (\mathbf{a}^T\boldsymbol{\sigma})(\mathbf{b}^T\boldsymbol{\sigma})(\mathbf{a}^T\boldsymbol{\sigma})=2(\mathbf{a}^T\mathbf{b})\mathbf{a}^T\boldsymbol{\sigma}-|\mathbf{a}|^2 \mathbf{b}^T\boldsymbol{\sigma}.
\end{gather}
This identity follows from applying the Pauli matrix equality $( \mathbf{a} \cdot \boldsymbol{\sigma} ) ( \mathbf{b} \cdot\boldsymbol{\sigma} )=( \mathbf{a} \cdot\mathbf{b} ) \mathbb{I}+i ( \mathbf{a} \times\mathbf{b} ) \boldsymbol{\sigma}$ twice. Setting $\mathbf{a}=\mathbf{v}_{i j^{( i )}}^{x_{i j^{( i )}}}$ and $\mathbf{b}=\mathbf{u}$, the second summand becomes 
\begin{gather}
    2\,\mathbf{u}^T \mathbf{v}_{i j^{( i )}}^{x_{i j^{( i )}}} {\mathbf{v}_{i j^{( i )}}^{x_{i j^{( i )}}}}^T\boldsymbol{\sigma}-\mathbf{u}^T \boldsymbol{\sigma},
\end{gather}
where we have used $\left| \mathbf{v}_{i j^{( i )}}^{x_{i j^{( i )}}}\right|=1$. Consequently, equation~\eqref{eq:coreqmiddlestep1} simplifies to
\begin{gather}\label{eq:23sum}
     \frac{1}{k}\sum_{\substack{a_{i j^{(i)}},\\
    x_{i j^{(i)}}}}
    \left\{ \frac{1-F_{{i j^{(i)}}}}{4}\mathrm{tr}\left[ 
    \left(
    \mathbf{u}^T \boldsymbol{\sigma}+2\,\mathbf{u}^T \mathbf{v}_{i j^{( i )}}^{x_{i j^{( i )}}} {\mathbf{v}_{i j^{( i )}}^{x_{i j^{( i )}}}}^T\boldsymbol{\sigma}-\mathbf{u}^T \boldsymbol{\sigma}\right)\rho_{i (j^{(i)}-1)} \right] \right\}
    =\frac{1-F_{i j^{(i)}}}{k}\mathrm{tr}\left[\mathbf{u}^T T_{i j^{(i)}}\boldsymbol{\sigma}\, \rho_{i (j^{(i)}-1)}\right].
\end{gather}
This result follows from summing over $a_{i j^{(i)}}$ and $x_{i j^{(i)}}$. The summation over all measurement settings $x_{i j^{(i)}}$ naturally gives rise to the measurement matrix $T_{i j^{(i)}} = \sum_{x_{i j^{(i)}}}\mathbf{v}_{i j^{( i )}}^{x_{i j^{( i )}}} {\mathbf{v}_{i j^{( i )}}^{x_{i j^{( i )}}}}^T$, which captures the collective contribution of all measurement directions. Equation~\eqref{eq:23sum} combined with~\eqref{eq:mainrecursiveR1} proves the claim.
\end{proof}

We now make several observations concerning~\eqref{eq:mainrecursiveR}. First, $\mathbf{u}$ need not be a unit vector. Second, $\mathbf{u}^T T_{i j^{(i)}}$ is itself a vector. The recursive structure is now evident: to obtain the final correlator, we apply~\eqref{eq:mainrecursiveR} recursively, starting from position $A_{i (j^{(i)}-1)}$ with measurement vector $\mathbf{v}_{i (j^{( i )}-1)}^{x_{i (j^{( i )}-1)}}$ and proceeding until $\rho_{i0}$. Returning to~\eqref{eq:JianhuiA3}, we apply the claim~\eqref{eq:mainrecursiveR} to the $\dfrac{1}{k}$-summation over $a_{i (j^{(i)}-1)}$ and $x_{i (j^{(i)}-1)}$ of $G_{i j^{(i)}}\mathrm{tr}\left[A_{ij^{(i)}}^{x_{ij^{(i)}}}\rho_{i (j^{(i)}-1)} \right]$, obtaining 
\begin{gather}
    G_{i j^{(i)}}\left\{ F_{i (j^{(i)}-1)} \, \mathrm{tr}\left[{\mathbf{v}_{i j^{( i )}}^{x_{i j^{( i )}}}}^T \boldsymbol{\sigma}      \rho_{i (j^{(i)}-2)}\right]\right.
    \left.
    +\frac{1}{k}\left[1- F_{i (j^{(i)}-1)} \right]\mathrm{tr}\left[ {\mathbf{v}_{i j^{( i )}}^{x_{i j^{( i )}}}}^T T_{i (j^{(i)}-1)}\boldsymbol{\sigma}\rho_{i (j^{(i)}-2)} \right]     
    \right\}\label{eq:mainrecursiveR2}.
\end{gather}
Alternatively, by introducing a binary variable $\varepsilon_q\in\{0,1\}$, we can write~\eqref{eq:mainrecursiveR2} more compactly as
\begin{gather}
    G_{i j^{(i)}}\sum_{\varepsilon_{ (j^{(i)}-1)}=0}^1 F_{i(j^{(i)}-1)}^{\varepsilon_{(j^{(i)}-1)}}\left[\dfrac{1-F_{i(j^{(i)}-1)}}{k} \right]^{1-\varepsilon_{(j^{(i)}-1)}}
    \mathrm{tr}\left\{ {\mathbf{v}_{i j^{( i )}}^{x_{i j^{( i )}}}}^T\left[T_{i(j^{(i)}-1)}\right]^{1-\varepsilon_{(j^{(i)}-1)}} \boldsymbol{\sigma}\,\rho_{i (j^{(i)}-2)} \right\}.\label{eq:Tj-1}
\end{gather}
Expressing~\eqref{eq:mainrecursiveR2} in binomial form streamlines our calculation, as we now have only one trace term to manipulate. Applying~\eqref{eq:mainrecursiveR} to~\eqref{eq:Tj-1}, with ${\mathbf{v}_{i j^{( i )}}^{x_{i j^{( i )}}}}^T\left[T_{i(j^{(i)}-1)}\right]^{1-\varepsilon_{(j^{(i)}-1)}}$ as our vector $\mathbf{u}$, the $\dfrac{1}{k}$-summation over $a_{i (j^{(i)}-2)}$ and $x_{i (j^{(i)}-2)}$ yields 
\begin{gather}
    G_{i j^{(i)}}\sum_{\varepsilon_{ (j^{(i)}-1)}=0}^1 F_{i(j^{(i)}-1)}^{\varepsilon_{(j^{(i)}-1)}}\left[\dfrac{1-F_{i(j^{(i)}-1)}}{k} \right]^{1-\varepsilon_{(j^{(i)}-1)}}\times
    \sum_{\varepsilon_{ (j^{(i)}-2)}=0}^1 F_{i(j^{(i)}-2)}^{\varepsilon_{(j^{(i)}-2)}}\left[\dfrac{1-F_{i(j^{(i)}-2)}}{k} \right]^{1-\varepsilon_{(j^{(i)}-2)}}\times\nonumber\\
    \mathrm{tr}\left\{ {\mathbf{v}_{i j^{( i )}}^{x_{i j^{( i )}}}}^T \left[T_{i(j^{(i)}-1)}\right]^{1-\varepsilon_{(j^{(i)}-1)}}\,\left[T_{i(j^{(i)}-2)}\right]^{1-\varepsilon_{(j^{(i)}-2)}}\boldsymbol{\sigma}\,\rho_{i (j^{(i)}-3)} \right\}.\label{eq:mainrecursiveReg}
\end{gather}
Proceeding recursively until we reach the initial density matrix $\rho_{i0}$, the trace term becomes
\begin{gather}
    \mathrm{tr}\left\{ {\mathbf{v}_{i j^{( i )}}^{x_{i j^{( i )}}}}^T \left[ \prod_{q=j^{(i)}-1}^{1} T_{i q}^{1-\varepsilon_q}\right] \, \boldsymbol{\sigma} \rho_{i0}\right\}
    =\frac{1}{4}\mathrm{tr}\left\{ {\mathbf{v}_{i j^{( i )}}^{x_{i j^{( i )}}}}^T \left[ \prod_{q=j^{(i)}-1}^{1} T_{i q}^{1-\varepsilon_q}\right] \, \boldsymbol{\sigma} \left( \mathbb{I} -b_i\,{\mathbf{w}_i^y}^T \boldsymbol{\sigma}  \right)   \right\}. \label{eq:mainrecursiveR3}
\end{gather}

Note that the product of the transposed vector and the ordered product of matrices is again a vector. Thus~\eqref{eq:mainrecursiveR3} can be evaluated as $\dfrac{1}{4}\mathrm{tr}[\mathbf{u}^T\boldsymbol{\sigma}(\mathbb{I}-b_i{\mathbf{w}_i^y}^T \boldsymbol{\sigma})]$, which equals $-\dfrac{b_i}{2}\mathbf{u}^T \mathbf{w}_i^y$. Substituting the result and summing over $b_i$ yields
\begin{gather}
    -\frac{G_{i j^{(i)}}}{2} \sum_{b_i\in\{-1,1\}}b_i^2\, {\mathbf{v}_{i j^{( i )}}^{x_{i j^{( i )}}}}^T \left[ \prod_{q=j^{( i )}-1}^{1} T_{i q}^{1-\varepsilon_q}\right]\mathbf{w}_i^y= -G_{i j^{(i)}}\,{\mathbf{v}_{i j^{( i )}}^{x_{i j^{( i )}}}}^T \left[ \prod_{q=j^{( i )}-1}^{1} T_{i q}^{1-\varepsilon_q}\right]\mathbf{w}_i^y.
\end{gather}

Collecting all successive summations, such as those over $\varepsilon_{(j^{(i)}-1)}$ and $\varepsilon_{(j^{(i)}-2)}$ in~\eqref{eq:mainrecursiveReg}, from $\varepsilon_{(j^{(i)}-1)}$ down to $\varepsilon_1$, we obtain for $j^{(i)}<m$:
\begin{gather}
    \langle{A}_{ij^{( i )}}^{x_{i j^{( i )}}} B_{i}^{y} \rangle=-\mathbf{v}_{i j^{( i )}}^{x_{i j^{( i )}} T} G_{i j^{(i)}}
    \left[ \sum_{\varepsilon_{1},\dots,\varepsilon_{(j^{(i)}-1)}=0}^1 \prod_{q=j^{( i )}-1}^{1} F_{iq}^{\varepsilon_q}\left(\dfrac{1-F_{iq}}{k} \right)^{1-\varepsilon_q}\left(T_{iq}\right)^{1-\varepsilon_q} \right] \mathbf{w}_{i}^{y}\, .\label{eq:mainrecursiveR4}
\end{gather}
For the case $j^{(i)}=m$, the final Alice performs a projective measurement, corresponding to $G_{i m}=1$. Moreover, $G_{i j^{(i)}}=\dfrac{1}{2}\left(\sum_{\varepsilon_{j^{(i)}}=0}^1 G_{i j^{(i)}}^{\varepsilon_{j^{(i)}}} G_{i j^{(i)}}^{1-\varepsilon_{j^{(i)}}} \right)$. We can thus absorb the $G_{i j^{(i)}}$ factor into the summands:
\begin{gather}
    \langle{A}_{ij^{( i )}}^{x_{i j^{( i )}}} B_{i}^{y} \rangle=-\mathbf{v}_{i j^{( i )}}^{x_{i j^{( i )}} T}
    \left[ \sum_{\varepsilon_{1},\dots,\varepsilon_{j^{(i)}}=0}^1 \prod_{q=j^{( i )}}^{1} \widetilde{K}_{iq} \right] \mathbf{w}_{i}^{y}\,
\end{gather}
where 
\begin{gather*}
    \widetilde{K}_{i q} =\left\{
    \begin{array}{l l l}
        \enspace\qquad\qquad\enspace \mathbb{I} &,\, q=j^{(i)}=m \\[0.5ex]
        \qquad\, \dfrac{1}{2} G_{i j^{(i)}}^{\varepsilon_{j^{(i)}}} G_{i j^{(i)}}^{1-\varepsilon_{j^{(i)}}}\mathbb{I} &,\, q =j^{(i)}<m \\[1.5ex]
         F_{iq}^{\varepsilon_q}\left(\dfrac{1-F_{iq}}{k} \right)^{1-\varepsilon_q}\left(T_{iq}\right)^{1-\varepsilon_q}\quad\quad &,  \, q<j^{(i)}\quad\quad.
    \end{array} \right.
\end{gather*}
Finally, we observe that the summations of multiplicative factors over binary sequences yield binomial expressions: $\sum_{\varepsilon=0}^1 x^\varepsilon y^{1-\varepsilon}=x+y$. We can therefore simplify $\tilde{K}_{i q}$ to obtain
\begin{equation}
    \frac{1}{2} \sum_{\varepsilon_{j^{(i)}}=0}^1 G_{i j^{(i)}}^{\varepsilon_{j^{(i)}}} G_{i j^{(i)}}^{1-\varepsilon_{j^{(i)}}} =G_{i j^{(i)}}
    \qquad \text{and}\qquad \sum_{\varepsilon_q =0}^1 F_{iq}^{\varepsilon_q}\left(\dfrac{1-F_{iq}}{k} \right)^{1-\varepsilon_q}\left(T_{iq}\right)^{1-\varepsilon_q}=F_{iq}\,\mathbb{I}+\frac{1-F_{iq}}{k} T_{iq}
\end{equation}
yielding the final expression
\begin{gather}
    \langle{A}_{ij^{( i )}}^{x_{i j^{( i )}}} B_{i}^{y} \rangle=-\mathbf{v}_{i j^{( i )}}^{x_{i j^{( i )}} T} \left[\prod_{q=j^{( i )}}^{1} K_{i q} \right] \mathbf{w}_{i}^{y}\,.
\end{gather}
where $K_{iq}$ is defined as~\eqref{eq:Kiqexpression}.

\twocolumngrid
\bibliography{Vertesi}
\end{document}